\documentstyle[12pt]{article}
\begin{document}
\begin{center}
{\Large{\bf Fractals at $T=T_c$ due to instanton-like configurations}}\\
\vspace{2 cm}
N.G. Antoniou, Y.F. Contoyiannis, F.K. Diakonos\\
\smallskip
{\it Department of Physics, University of Athens, GR-15771 Athens, Greece}\\
\vspace{1.2 cm}
and\\
\vspace{0.5cm}
C.G. Papadopoulos \\
{\it NRCPS, `Demokritos', Aghia Paraskevi, GR-15310 Athens, Greece}\\
\end{center}
\vspace{2.8 cm}
\begin{abstract}
We investigate the geometry of the critical fluctuations for a general
system undergoing a thermal second order phase transition. Adopting a
generalized effective action for the local description of the fluctuations of the
order parameter at the critical point ($T=T_c$) we show that instanton-like
configurations, corresponding to the minima of the effective action
functional,
build up clusters with fractal geometry characterizing locally the critical
fluctuations. The connection between the corresponding
(local) fractal dimension and the critical exponents is derived. Possible
extension of the local geometry of the system to a global picture is also
discussed.
\end{abstract}

\newpage
In a system undergoing a second order phase transition, coherent
fluctuations at all scales, occur, at the critical
point. The understanding of the geometry
of these fluctuations is a long-standing problem \cite{FisBi}. The
self-similarity of the critical fluctuations suggests the formation of
clusters, with nonvanishing value of the order parameter, which have a
fractal structure \cite{Mand,Stin}. Simple geometrical arguments lead to the
conclusion that the fractal dimension, describing the critical clusters,
reflects the scaling properties of the underlying fluctuations and therefore
can be related to the critical exponents characterizing the phase transition
\cite{Stin}. Rigorous mathematical derivation and a deeper understanding of
the origin of such a relation is however missing in the general case.
Some important efforts have been performed concerning the geometry of the
critical clusters in the Ising and Potts models \cite{Coni} where the order
parameter is described through a discrete field variable.
For the more general case of continuous fields there is however, to our
knowledge, no such understanding. It is the purpose of the present paper to
illuminate the way the geometry of the critical clusters emerges in the case
of a continuous effective field theory.
For our considerations we
study a self-interacting scalar field $\phi$ at thermal equilibrium.
The effective action of the thermal system, at the critical point $T=T_c$ of
the continuous phase transition, can be specified, in a wide class of critical
phenomena, by an effective theory in $d$ dimensions, in terms of a
macroscopic field $\phi$ (order parameter) as follows:
\begin{equation}
\Gamma_c[\phi]=g_1 \Lambda^{-d-2}
\int d^d x \left[\frac{1}{2}(\nabla_d \phi)^2 +
g_2 \Lambda^{2 d+2} \vert \Lambda^{-d} \phi \vert^{\delta+1} \right]
\label{fener}
\end{equation}
The dimension of the field $\phi$ in (\ref{fener}) has been chosen:
$\phi \sim$ (volume)$^{-1}$ and the ultraviolet cut-off $\Lambda$ of the
underlying microscopic theory fixes the coarse graining scale $R_c \approx
\Lambda^{-1}$ of the effective system (throughout this work we use the
convention $\kappa_B=1$ (Boltzman constant) and the energy is given in
inverse length units). The form (\ref{fener}) for constant
fields leads to the standard equation of state at $T=T_c$:
$\frac{\delta \Gamma_c}{\delta \phi} \sim \phi^{\delta}$ $(\phi > 0)$
and therefore the index $\delta$
is identified with the isothermal critical exponent of the system. For a
high-temperature phase transition, the coarse graining cutoff $\Lambda$
is bounded by the critical temperature itself $(\Lambda\gg T_c)$ and the
dimensionless parameters $g_1$, $g_2$ in eq.(\ref{fener}) are expressed
in terms of the ratio $\lambda=\frac{\Lambda}{T_c}$ $(\lambda \gg 1)$.
In fact, using as a concrete example the $O(N)$ $3d$ effective theory, the
action $ \Gamma_c[\phi]$ in the large $N$ limit and for a fixed
orientation in the internal $O(N)$ space, is written as follows \cite{Wett}:
$$ \Gamma_c[\phi]= \lambda^5 \Lambda^{-5}
\int d^3 x [\frac{1}{2} (\nabla \phi)^2 + 2
\left(\frac{2 \pi \lambda^5}{N} \right)^2 \Lambda^8
(\Lambda^{-3} \phi)^6 ] $$
It belongs to the general class (\ref{fener}) with:
$d=3,~\delta=5,~g_1=\lambda^5,~g_2=2\left(\frac{2 \pi \lambda^5}{N}\right)^2$
and to the particular sector $g_1 \gg 1$.

The scalar field $\phi$ in the following will be no further
specified. It can describe magnetization density or particle density or
the density of any other extensive physical quantity characterizing the
phase transition. Introducing the dimensionless quantities:
$\hat{\phi}=\Lambda^{-d} \phi~~~;~~~\hat{x}_i=\Lambda x_i$
we can rewrite the effective action (\ref{fener}) as:
\begin{equation}
\Gamma_c[\hat{\phi}]=g_1 \int_V d^d \hat{x}
\left[\frac{1}{2}(\nabla \hat{\phi})^2 +
g_2 \vert \hat{\phi} \vert^{\delta+1} \right]
\label{feners}
\end{equation}
In what follows we use, for simplicity, the old notation $(\phi,x_i)$ instead
of $(\hat{\phi},\hat{x}_i)$.

The statistics of the critical system is resolved if we are able
to calculate the partition function:
\begin{equation}
Z=\int {\cal{D}}[\phi] e^{-\Gamma_c[\phi]}
\label{parf}
\end{equation}
The nontrivial task is to carry out the path integration in (\ref{parf}).
Since the self-similarity is a subtle symmetry of the system and it is
expected to dominate in the formation of the fractal geometry of the
clusters (the precise definition of the cluster will be given later on) the
conventional methods to perform the path integration, which can be found in
the literature \cite{Scal}, are not suitable for the problem at hand. We
propose therefore to perform the summation, with an appropriate measure,
over a class of saddle point configurations which are expected to dominate
the critical fluctuations in eq.(\ref{feners}) for $g_1 \gg 1$.
In order to illustrate our
method we will treat, for simplicity, the one-dimensional case in some
detail. Our approach can be however extended to higher dimensions without
difficulties.

Let us now become more quantitative. In the one-dimensional case the
partition function of the critical system ($T=T_c$) is given as:
\begin{equation}
Z=\int {\cal{D}}[\phi] \exp \left[ -g_1 \int_0^R dx
\left[ \frac{1}{2} \left(\frac{d \phi}{dx}\right)^2
+ g_2 \vert \phi \vert^{\delta+1} \right] \right]
\label{parf1d}
\end{equation}
where $R$ is the size of the considered system. We assume here that our
investigations refer to an open subsystem of the entire physical system
located in the vicinity of the point $x=0$.
Therefore no
restrictions to the values of the order parameter at the boundaries of the
subsystem are imposed.
The geometrical properties of
the subsystem $R$ are expressed through the scaling properties of
the extensive quantities characterizing the subsystem with
varying size $R$ around $x=0$.

The saddle-point configurations $\phi(x)$ fullfill the Euler-Lagrange
equation corresponding to the effective action $\Gamma_c[\phi]$ and describe
the classical motion in the concave potential
$U(\phi) \sim -\vert \phi \vert^{\delta +1}$.
This equation 
can be solved analytically in terms of two parameters
$E$ and $\phi(0)$, where $E$ is a conserved (during the classical motion)
quantity identified with the total energy of the moving particle:
\begin{equation}
E=\frac{1}{2}\left(\frac{d \phi}{dx}\right)^2-g_2 \vert \phi
\vert^{\delta +1}
\label{cener}
\end{equation}
Using eq.(\ref{cener}) one can show that configurations with $E \neq 0$
contribute to the partition function $Z$ with a suppresion factor
$e^{- g_1 R \vert E \vert}$ suggesting that the dominant saddle
points contributing to (\ref{parf1d})
come from those solutions of the equations of motion for
which $E \approx 0$. In fact eq.(\ref{cener}) can be integrated to give,
for $E=0$, instanton-like solutions of the form:
\begin{equation}
\phi(x)=\left( \frac{c}{\sqrt{2 g_2}} \right)^{\frac{2}{\delta -1}}
\left[ \frac{c}{\sqrt{2 g_2}} \phi(0)^{-\frac{\delta -1}{2}} \pm x
\right]^{-\frac{2}{\delta -1}} ~~~~;~~~
(\delta > 1)
\label{ansol}
\end{equation}
with $c=\frac{2}{\delta -1}$.
Setting $x_o=\mp \frac{c}{\sqrt{2 g_2}} \phi(0)^{-\frac{\delta -1}{2}}$ the
$\phi$-field for $E=0$ simplifies to:
\begin{equation}
\phi(x)=A \vert x - x_o \vert^{-\frac{2}{\delta - 1}}~~~~~~;~~~~~~
A=\left[ \frac{g_2}{2} (\delta -1)^2 \right]^{-\frac{1}{\delta -1}}
\label{classo}
\end{equation}
To perform the path integration in (\ref{parf1d}) we have to sum up the
contributions of all instanton-like saddle point configurations of the
form (\ref{classo}), i.e. to integrate over the parameter $x_o$. In order to
determine the correct integration measure 
we consider the class of solutions (\ref{classo}) with $x_o \gg R$. In this
case $\phi(x)=$constant $\sim x_o^{-\frac{2}{\delta -1}}$ and in a region
of radius $R$ around $x=0$ the path integration becomes an ordinary
integral over $x_o$ with measure: ${\cal{D}} \phi=d \mu (x_o) \approx
x_o^{-\frac{\delta +1}{\delta -1}} d x_o$. To determine the range of
integration over $x_o$ we have first to clarify the meaning of a cluster
in our picture: Let us assume that $M$ is an extensive variable
(i.e. magnetization)
characterizing the field configurations of the critical system and possesing
a minimal, in general different from zero, value $\mu$ ($M \geq \mu$)
related to the microscopic details of the system.  
Within the
picture of the local observer positioned at $x=0$ a cluster of size $R$
is the set $S$ of points with a maximum distance $R$ from the
origin.
The
most appropriate observable to study the geometric properties of the critical
clusters, is the thermal average $<M(R)>=< \int_0^R \phi(x) dx >$
and in particular its behaviour as a function of $R$.
The minimum value $\mu$ of the magnetization $M$ introduces a
threshold $\phi_{min}$ to the configurations $\phi$ contributing to this
average, leading us to an upper limit for the integral $x_o$ as a function
of $R$: $x_o \leq \left(\frac{A R}{\mu}\right)^{\frac{\delta -1}{2}}$.
On the other hand the singularity at $x=x_o$ must
lie outside the region of the considered cluster $(0,R)$ restricting
the integration in (\ref{parf1d}) over configurations with $x_o~\geq~R$.
In terms of the partition function
(\ref{parf1d}) this average can be determined as:
\begin{eqnarray}
<\int_0^R \phi(x) dx>&=&\frac{A \frac{\delta-1}{\delta-3}}{Z}
\displaystyle{\int_R^{(A R/ \mu)^{\frac{\delta -1}{2}}}}
d x_o \cdot x_o^{-\frac{\delta +1}
{\delta-1}}[x_o^{\frac{\delta -3}{\delta -1}}-(x_o - R)^
{\frac{\delta -3}{\delta -1}}] \\
&\cdot&\exp[- G_1 \left( \frac{\delta -1}{\delta +3} \right)
[ (x_o - R)^{-\frac{\delta +3}{\delta -1}}-
x_o^{-\frac{\delta +3}{\delta -1}}]] \nonumber
\label{avfac}
\end{eqnarray}
with $G_1=2 g_1 g_2 A^{\delta +1}$.
One can show analytically that in the large $G_1$ limit ($G_1 \gg 1$) there
are three characteristic regions determining the behaviour of the integral
in eq.(8). Putting
$R_d=A^{-\frac{\delta +1}{\delta}} \mu^{\frac{\delta +1}{\delta}}
G_1^{\frac{1}{\delta}}$ and
$R_u = G_1^{\frac{\delta -1}{\delta +3}}$ we find that
in the central region of scales $R_d \ll R \ll R_u$:
$<\int_0^R \phi(x) dx > \sim R^{\frac{\delta}{\delta + 1}}$
leading to a fractal structure of the
cluster around the point $x=0$ with a fractal mass dimension \cite{Meak}:
$$d_F=\frac{\delta}{\delta +1}$$
This behaviour crosses over for $R \gg R_u$ to a different
power-law:
$$<\int_0^R \phi(x) dx > \sim R^{\frac{\delta -3}{\delta -1}}$$
suggesting the presence of a fractal with mass dimension $\tilde{d}_F=
\frac{\delta -3}{\delta -1}$ at large scales.

For $R \ll R_d$ a violation of the scaling symmetry of the critical cluster
 is revealed
leading to an approximately constant value of the integral
(8). The parameter $R_d$ defines a minimal scale of the critical
system effectively related with the minimal value $\mu$ of the order
parameter.
In Fig.1a we show the numerical results for the calculation of
(8) using the values $G_1=5 \cdot 10^8$ and $\delta=5$.
The three different regions and the corresponding cross-over scales
describing the geometry of the critical cluster 
are clearly distinguished. In the same plot we show also the corresponding
linear fits to illustrate more transparently the above considerations.
The fractality in the central region characterizes the critical system in the
sense that it corresponds to the scaling behaviour in the vicinity of the
local observer when $\mu \to 0$.
This is at best shown in Fig.1b where we calculated (8) using
the same value of $G_1$ as in Fig.1a and let the upper limit in the
$x_o$-integration going to infinity. The crossover scale $R_u$ gives
presumably a measure of the correlation length of the finite system at
$T=T_c$.

One can now easily generalize the above
investigations in order to describe systems of higher dimensions $(d~>~1)$.
We proceed in a similar
way as for the one-dimensional case using the saddle point approximation
for the partition function $Z_d$:
\begin{equation}
Z_d=\int {\cal{D}}[ \phi ] e^{- g_1 \int_V d^d x
[\frac{1}{2} \left({\nabla_d \phi} \right)^2
+ g_2 \vert \phi \vert^{\delta +1}]}
\label{parfmd}
\end{equation}
in order to calculate the thermal average
$< \int_0^R \phi(\vec{x})~d^d x >_{Z_d}$.
The summation over the saddle points in eq.(\ref{parfmd}) becomes an ordinary
integration over the position of the
singularity in the instanton-like solutions $\phi_d$ in $d$-dimensions.
In an analogous way as in the $1d$ case, solving the Euler-Lagrange
equations for the critical action, we get instanton-like solutions for
$\phi_d$. In the case $d=2$ a class of analytic solutions possessing a
point singularity can be determined:
\begin{equation}
d=2~~~;~~~~\phi_2(\vec{r})=A_2
\vert \vec{r} - \vec{r}_o \vert^{\frac{-2}{\delta-1}}~~~;~~~~
A_2=\left( \frac{g_2}{2}(\delta -1)^2 \right)^{\frac{-1}{\delta -1}}
\label{sol2d}
\end{equation}
Performing the calculation of the mean value :
$<M(R)>=<\int d^2 \vec{r} \phi(\vec{r})>$,
characterizing a two-dimensional critical cluster,
in an analogous way as for the $1d$ case,
we get a similar behaviour concerning its fractal geometric
properties. There are characteristic scales in the radial component
$R_d=A_2^{- \frac{\delta +1}{2 \delta }}
\mu^{\frac{\delta +1}{2 \delta}} G_2^{\frac{1}{2 \delta }}$ and
$R_u=G_2^{\frac{\delta -1}{4}}$ with $G_2=\pi g_1 [
\frac{2 A_2^2}{(\delta -1)^2} + g_2 A_2^{\delta + 1} ]$ such that:
\begin{eqnarray}
d=2~~~~;~~~~<M(R)> &\sim& R^{\frac{2 \delta}{\delta +1}}~~~~;~~~~~
R_d \ll R \ll R_u \nonumber \\
<M(R)> &\sim& R^{\frac{2 (\delta -2)}{\delta -1}}~~~~;~~~~~R_u \ll R
\label{pow2d}
\end{eqnarray}
A cross over for large $R$ is found also in this case. For dimensions
$d \geq 3$ no analytic solution to the Euler-Lagrange equations is available
in the general case of a non vanishing anomalous dimension $\eta$
($\delta=\frac{d +2 - \eta}{d - 2 + \eta}$ \cite{Stan}).
However these equations can be integrated numerically
leading again to an instanton-like behaviour. In particular one can find
exact analytic spherical solutions of this kind for $d \geq 3$ in the special
case when $\eta=0$ $(\delta=\frac{d +2}{d -2})$.

For $0 < \eta \ll 1$ an approximate solution can be obtained
given as follows:
\begin{equation}
d \geq 3~~~;~~~~\phi_d(r)=A_d (r^2_o - r^2)^{\frac{2-d}{2}}~~~;~~~~
A_d=\left(\frac{(d-2) r_o}{\sqrt{2 g_2}} \right)^{\frac{d-2}{2}}
\left(\frac{(d-2)}{\sqrt{2 g_2} r_o} \right)^{\frac{d \eta}{4}}
\label{sol3d}
\end{equation}
The solution (\ref{sol3d}) goes to the exact one for $\eta=0$.
For $r$ far from the singularity region the approximate form (\ref{sol3d})
coincides practically with the exact (numerically obtained) solution.
This can be at best seen in Fig.2a where we plot together the numerical and
the approximate solution to the Euler-Lagrange equations for $d=3$ and
$\eta=0.34$. In fact, in a wide range of universality classes including the
$O(4)$ theory in which $\eta \approx 0.034$ \cite{Tetr}, the
anomalous dimension for $d=3$ is much smaller \cite{Ma}
and therefore one can safely use the solution
(\ref{sol3d}) for most calculations. Based on (\ref{sol3d}) we
determined $<M(R)>$ for spherically symmetric clusters in $d \geq 3$
dimensions and explored the geometric properties of such a cluster.
Once again we got the typical central fractality region crossing over
to a fractal with a smaller dimension for distances comparable to
the correlation length. It must be noted that for $d \geq 3$
the cross over disapears as
$\eta \to 0$. Using $R_d=a^{\frac{-(\delta + 1)}{d \delta}}
\mu^{\frac{\delta +1}{d \delta}}
(G_d)^{\frac{1}{d \delta}}$ and $R_u=(G_d)^{\frac{-1}{d + q (\delta +1)}}$
with $a=\left(\frac{d-2}{\sqrt{2 g_2}}\right)^{\frac{d-2 + \frac{d \eta}{2}}
{2}}$, $G_d=\frac{2 a^{\delta + 1} \pi^{d/2}}{d \Gamma(d/2)} g_1 g_2$ and
$q=\frac{2-d-\frac{d \eta}{2}}{2}$, we obtain:
\begin{eqnarray}
d \geq 3 ~~~~;~~~~~<M_d(R)> &\sim& R^{1+\frac{d - \eta}{2}}~~~~~;~~~~~
R_d \ll R \ll R_u \nonumber \\
<M_d(R)> &\sim& R^{1+\frac{d (2- \eta)}{4}}~~~~~;~~~~~R_u \ll R
\label{pow3d}
\end{eqnarray}
The characteristic behaviour of $<M(R)>$ for $d=3$ is presented in Fig.2b.
Here we used $G_3=10^2$ and $\eta=0.34$ (as in Fig.2a). The breaking of the
fractality (for $R \ll R_d$) is clearly reproduced while the cross over is
suppressed due to the small value of $\eta$.
The power-laws  $<M_d(R)> \sim R^{d_F}$ or $\sim R^{\tilde{d}_F}$
with $d=1,2,..$ determine fractals at different scales
with dimensions $d_F, \tilde{d}_F$. Putting together our results in 1,2 and 3
dimensions and taking into account that our considerations for $d \geq 3$
are restricted to the case when the anomalous dimension $\eta$ is small
($\eta \ll 1$) we can cast our results for the fractal properties of the
critical cluster into universal expressions determining
$d_F$ and $\tilde{d}_F$ in terms of $d$ and $\delta$:
\begin{equation}
d_F=\frac{d \delta}{\delta + 1}~~~~;~~~~\tilde{d}_F=d - \frac{2}{\delta -1}
~~~~;~~~~d_F-\tilde{d}_F=\frac{\eta (d -2+ \eta)}{2 (2-\eta)}
\label{frac}
\end{equation}
The expression for $d_F$ is in accordance with the results obtained in
\cite{Stin,Coni} for the Ising and Pott's critical clusters.
Thus we have found that eqs.(\ref{frac})
describe the fractal geometry of the critical clusters in a wide
range of scales and for a general class of effective theories.
Considering the asymptotic region where the size of the cluster reaches the
value of the correlation length of the finite system we find a cross over to
a more dilute phase with a smaller fractal dimension $\tilde{d}_F$.
Furthermore we
have revealed a mechanism responsible for the formation of the geometry
of the critical clusters.
Our considerations are restricted to the point of view of a local
description.
The generalization of our approach in order to build up the entire critical
system, requires the extension of the formalism to configurations
incorporating many, suitably located, instanton-like structures
(of the size of the correlation length) covering the whole available space.
Such an investigation however goes beyond the scope of the present letter. \\

\noindent {\bf Acknowledgments} \\
We thank Dr. Ofer Biham for his helpful comments after reading the paper.
This  work was supported in part by contracts with the Hellenic General
Secretariat for Research and Technology ($\Pi ENE \Delta$ 1177, 8875/28-8-96)
and with the European Community (ERBFMBICT961541).

{} 

\vspace*{1.0cm}

\begin{center}
{\Large \bf Figure captions}
\end{center}

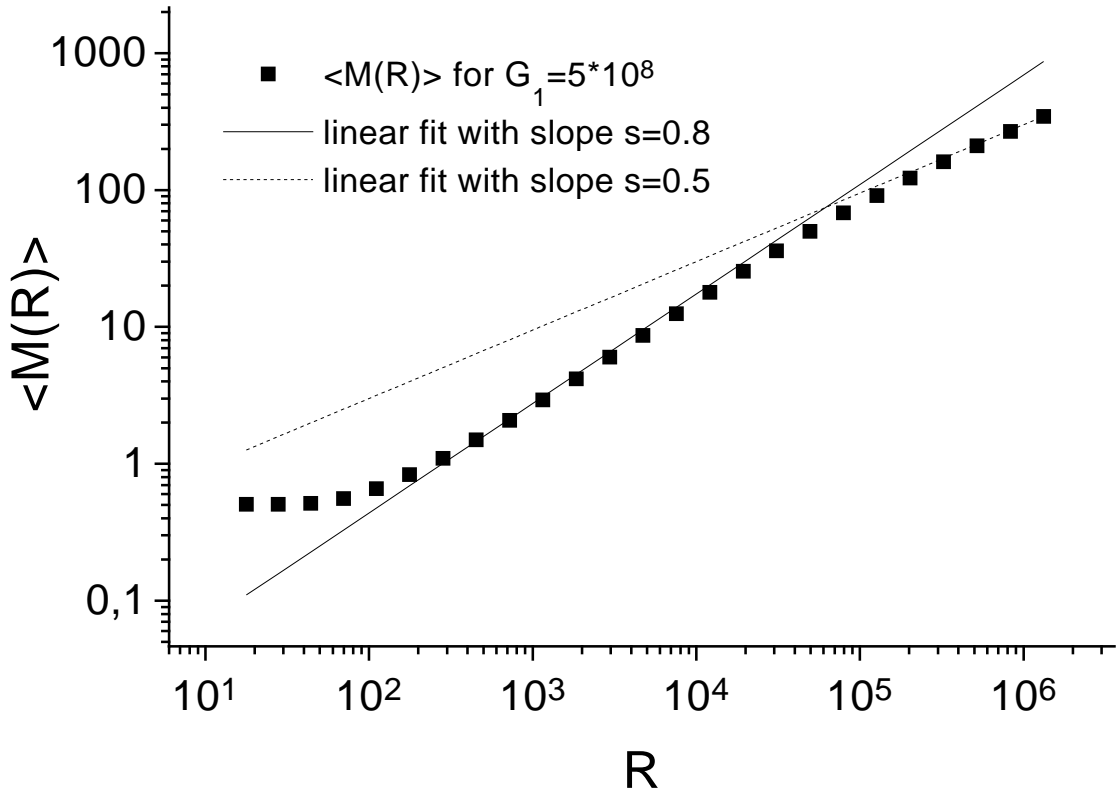
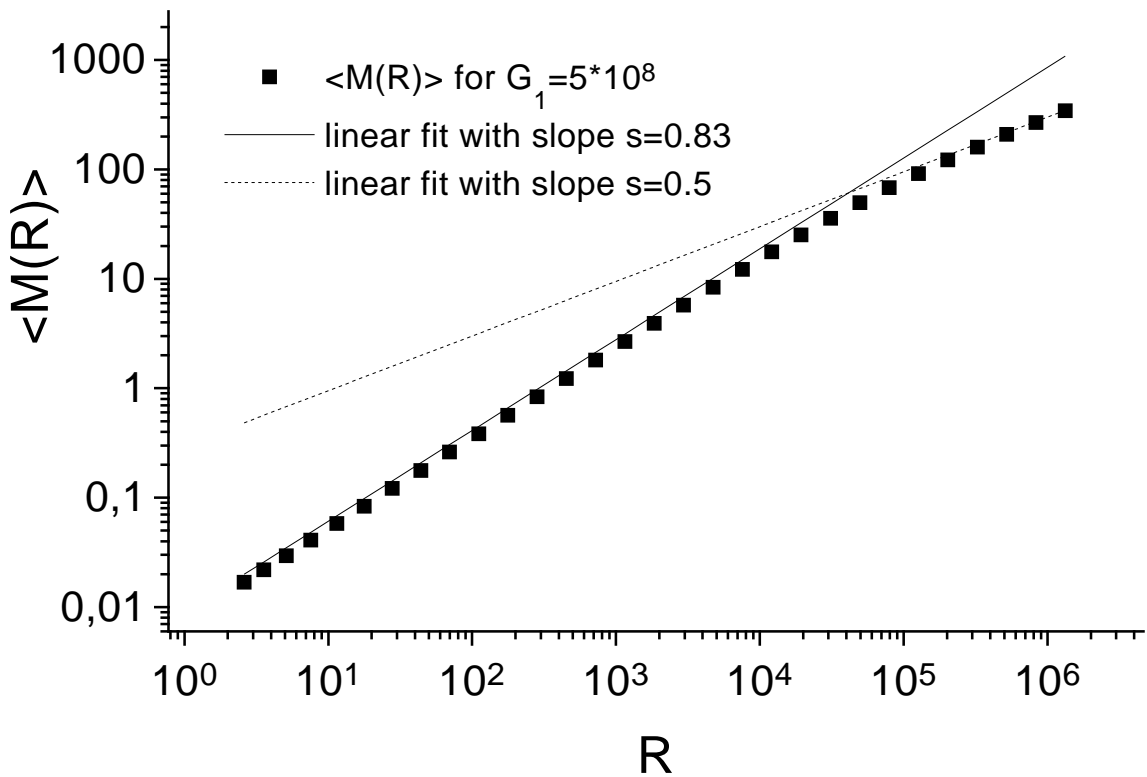
\noindent Figure 1: (a) The mean magnetization $<M(R)>$ as a function of $R$
(in units $\Lambda^{-1}$) for the $d=1$ case, $G_1=5 \cdot 10^8$ and $A=1$.
The fitted lines indicate the two regions of fractality as
described in the text. A nonvanishing minimum magnetization $\mu$ takes care
for the violation of the scaling in small distances $R$. \\
(b) The mean magnetization $<M(R)>$ for $d=1$, $G_1=5 \cdot 10^8$ and $A=1$
for $\mu \to 0$.

\vspace*{0.4cm}
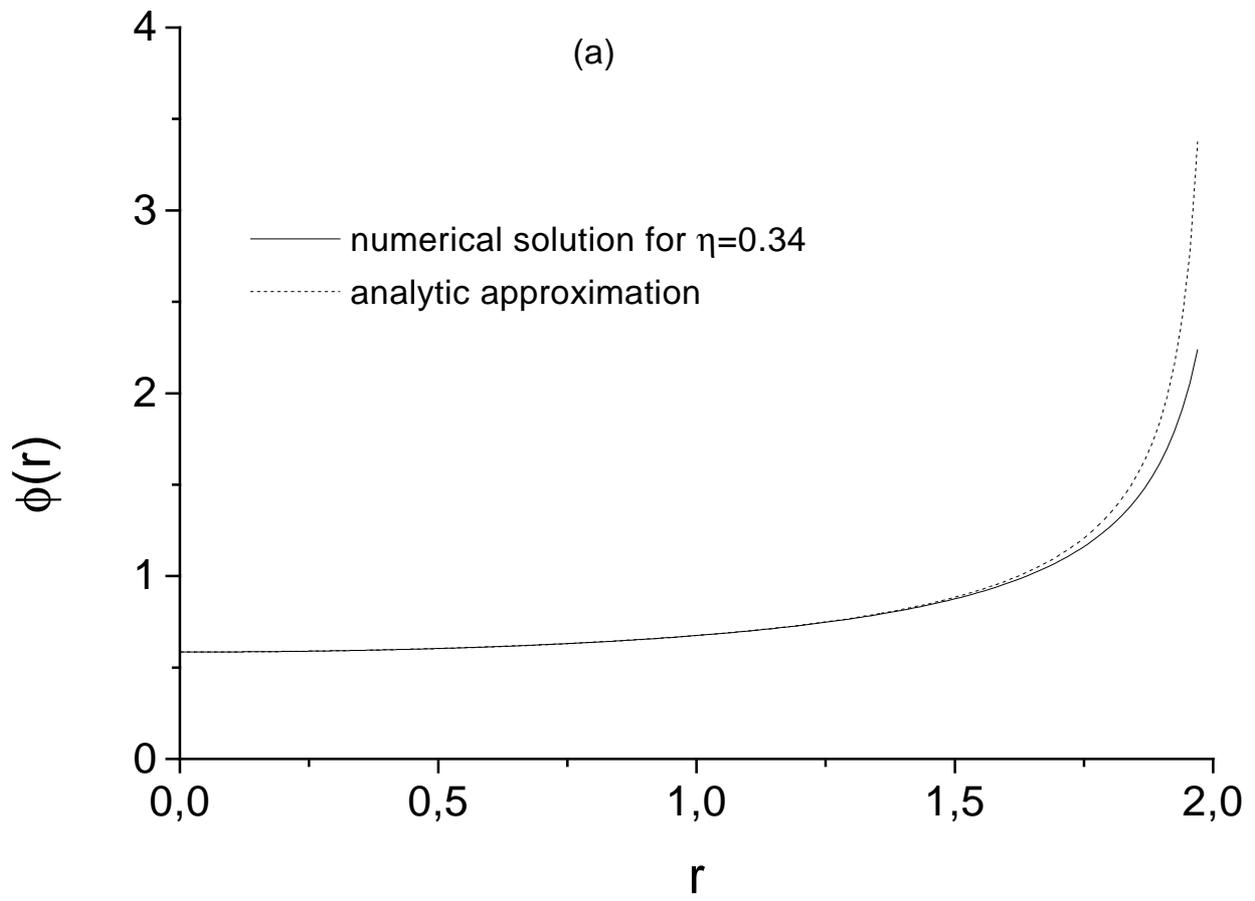
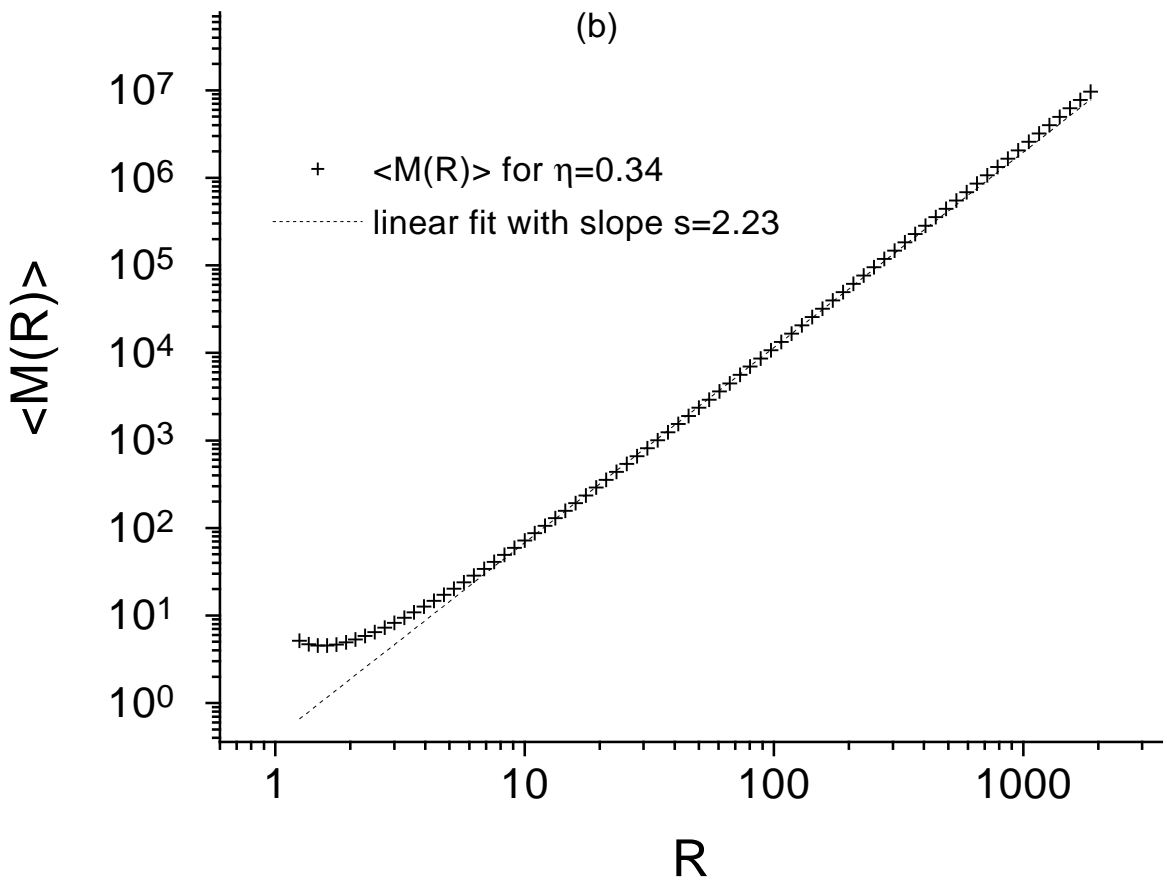
\noindent Figure 2: (a) A saddle point solution to the action (\ref{fener})
for $d=3$ and $\eta=0.34$. The analytical approximation (dashed line) and
the result of the nummerical integration (solid line) are displayed
separately. \\
(b) The mean magnetization $<M(R)>$ for $d=3$ and $G_3=10^2$. The fitted line
indicates the fractality in the central region.
\end{document}